\pdfoutput=1
\RequirePackage{fix-cm}
\documentclass[smallextended]{svjour3}       
\usepackage{graphicx}
\usepackage{mathtools}
\usepackage{amsfonts}
\usepackage{amssymb}
\usepackage{amsmath}
\makeatletter
\usepackage{colortbl}
\usepackage{wrapfig}
\usepackage{booktabs}
\usepackage{subcaption}
\def\longleftrightarrowfill@{\arrowfill@\leftarrow\relbar\rightarrow}
\makeatother
\usepackage[table]{xcolor}
\usepackage{adjustbox}
\usepackage{multirow}

\usepackage[utf8]{inputenc}
\usepackage{fourier} 
\usepackage{array}
\usepackage{makecell}

\usepackage{calrsfs}
\DeclareMathAlphabet{\pazocal}{OMS}{zplm}{m}{n}
\SetMathAlphabet\pazocal{bold}{OMS}{zplm}{bx}{n}
\usepackage{graphicx}
\usepackage{grffile}
\begin{document}

\title{Graph Classification via Heat Diffusion on Simplicial Complexes
}


\author{Mehmet Emin Aktas         \and
        Esra Akbas 
}


\institute{M. Aktas \at
              University of Central Oklahoma \\
              Department of Mathematics and Statistics\\
              \email{maktas@uco.edu}           
           \and
           E. Akbas \at
              Oklahoma State University \\
              Department of Computer Science \\
              \email{eakbas@okstate.edu}
}


\maketitle

\begin{abstract}
In this paper, we study the graph classification problem in vertex-labeled graphs. Our main goal is to classify the graphs comparing their higher-order structures thanks to heat diffusion on their simplices. We first represent vertex-labeled graphs as simplex-weighted super-graphs. We then define the diffusion Fr{\'e}chet function over their simplices to encode the higher-order network topology and finally reach our goal by combining the function values with machine learning algorithms. Our experiments on real-world bioinformatics networks show that using diffusion Fr{\'e}chet function on simplices is promising in graph classification and more effective than the baseline methods. To the best of our knowledge, this paper is the first paper in the literature using heat diffusion on higher-dimensional simplices in a graph mining problem. We believe that our method can be extended to different graph mining domains, not only the graph classification problem.

\end{abstract}

\section{Introduction}
Graphs (networks) are important structures used to model complex data where nodes (vertices) represent entities and edges represent the interactions or relationships among them \cite{aggarwal2010managing,Cook2006}. We can see the applications of network data in many different areas such as (1) social networks consisting of individuals and their interconnections such as Facebook, coauthorship and citation networks of scientists~\cite{akbas2017attributed,akbas2017truss,tanner2019paper}, (2) protein interaction networks from biological networks consisting of proteins that work together to perform some particular biological functions~\cite{Newman06062006,przulj2003graph}. Furthermore, we can assign \textit{labels} to the graph structures such as vertices, edges, or whole graphs.  As an example to a vertex-labeled network, in a World Wide Web network of a university, where webpages are vertices and hyperlinks are edges, we can label the nodes as faculty, course, department. As another example, in chemoinformatics, molecules are represented as labeled networks depending on their properties such as anti-cancer activity, molecule toxicity. In addition to labels of each network, vertices, representing atoms, can have labels based on groups of atoms.

Graph classification, as one of many network analysis applications, is the task of identifying the labels of graphs in a graph dataset. It found applications in many different disciplines such as biology \cite{borgwardt2005protein}, chemistry \cite{duvenaud2015convolutional}, social network analysis \cite{keil2018topological} and urban planning \cite{bao2017planning}. For example, in chemistry, the graph classification task can be detecting labels of graphs, for instance, anti-cancer activity or molecule toxicity.

Although graph classification is practical and essential, there are some challenges of using machine learning algorithms in this task:

\begin{enumerate}
    \item The enormous sizes of real-world networks make the existing solutions for different graph classification problems hard to adapt with the high computation and space costs.
    
    \item Most current methods only use the vertex and edge information in networks, i.e., only pairwise relations between entities, for classification. However, as we see in different real-world applications, such as human communication, chemical reactions, and ecological systems, interactions can occur in groups of three or more nodes. They cannot be simply described as pairwise relations~\cite{battiston2020networks}.
    
    \item  Graph data is complex, and its non-linear structure makes it difficult to apply machine learning algorithms. For example, one cannot use regular measures to compute the similarity between two graphs. These graphs may have a different number of vertices and edges with no clear correspondence.
\end{enumerate} 

In this paper, we study the graph classification problem by modeling the higher-order interaction among graphs via heat diffusion on simplicial complexes. While our primary goal in this paper is to address the second challenge of the graph classification using higher-dimensional simplicial complexes, our novel approach also addresses the other challenges as well. Our graph classification model employs the simplex-weighted super-graph representation, heat diffusion on simplicial complexes, and machine learning techniques. We first represent a vertex-labeled graph as a simplex-weighted super-graph with super-nodes being the unique vertex labels. This step compresses the original large graph into a relatively small graph, keeping the crucial information of the original graph as simplex weights in the compressed graph. This scales up the graph classification, hence, addresses the first challenge. Then, we define the heat diffusion not only on vertices (i.e., 0-simplices) but also on higher dimensional simplices such as edges, triangles (i.e., 1-simplices, 2-simplices). We further design the diffusion Fr{\'e}chet function (DFF) on simplices of the network to extract the higher-order graph topology. DFF is the right choice since it is defined based on the topological and geometrical structure of higher-order graph architecture thanks to the heat diffusion. That is why it allows us to take the higher-order structures in graphs into consideration for the classification problem, hence addresses the second challenge. Applying DFF on a simplex-weighted super-graph will give a value to each super-node in the graph. Then, we create a uniform feature list using the labels of super-nodes for each graph and use DFF values of super-nodes as its features. This feature list provides a clear correspondence between graphs, hence addresses the third challenge. As the last step, we employ machine learning algorithms for classification using these features. We summarize our contributions as follows.

\begin{itemize}
    \item We design simplex-weighted super-graphs with super-nodes being the unique vertex labels in the vertex-labeled graph. This relatively small graph scales up the graph classification
    \item We define the heat diffusion on higher dimensional simplices such as edges, triangles and develop the diffusion Fr{\'e}chet function (DFF) for higher-dimensional simplices to capture the complex structure of the higher-order graph architecture. 
    \item We represent each graph as a vector using the nodes' labels and their DFF values and use these features for learning.
\end{itemize}

The paper is structured as follows. In Section \ref{sec:prelim}, we give the necessary background for our method. We first give a formal definition to a network and the graph classification problem, then define simplicial complexes, simplicial Laplacian, and the diffusion Fr{\'e}chet function on manifolds. We also provide related work in this section. In Section \ref{sec:method}, we introduce our graph classification model with explaining our simplex-weighted super-graphs and diffusion Fr{\'e}chet function on simplicial complexes. In Section \ref{sec:exp}, we present our experimental results and compare them with the baseline methods. Our final remarks are reported in Section \ref{sec:conc}.

\section{Background}\label{sec:prelim}
In this section, we discuss the preliminary concepts for networks, graph classification problem, simplicial complexes, and the diffusion Fr{\'e}chet function (DFF). We also elaborate on related work with a particular focus on the graph classification problem and DFF in graph mining.
\subsection{Preliminaries}

\subsubsection{Networks}
In a formal definition, a network $G$ is a pair of sets $G = (V, E)$ where $V$ is the set of vertices and  $E \subset V \times V $ is the set of edges of the network. There are various types of networks that represent the differences in the relations between vertices. While in an \textit{undirected network}, edges link two vertices symmetrically, in a \textit{directed network}, edges link two vertices asymmetrically. If there is a score for the relationship between vertices that could represent the strength of interaction, it is represented as a \textit{weighted network}. In a weighted network, a weight function $W : E \rightarrow \mathbb{R}$ on edges is defined to assign a weight for each edge. Furthermore, If vertices of a network have labels, we call these networks as \textit{vertex-labeled network}. More formally, for a vertex-labeled network, there is a function $l : V \rightarrow L$ defined on vertices that assign a label from the label set $L$ to each vertex.

Graph classification problem is the task of classifying graphs into categories. More formally, given a set of graphs $\mathbb{G}$ and a set of class labels $L$, the graph classification task is to learn a model that maps graphs in $\mathbb{G}$ to the label set $L$. 

\subsubsection{Simplicial complexes}
A \textit{simplicial complex} $K$ is a finite collection of simplices, i.e., points, edges, triangles, tetrahedron, and higher-dimensional polytopes, such that every face of a simplex of $K$ belongs to $K$ and the intersection of any two simplices of $K$ is a common face of both of them. 0-simplices correspond to vertices, 1-simplices to edges, 2-simplices to triangles, and so on.  

Let $S_p(K)$ be the set of all $p$-simplices of $K$. An $i$-\textit{chain} of a simplicial complex $K$ over the field of integers is a formal sum of its $i$-simplices and $i$-th \textit{chain group} of $K$ with integer coefficients, $C_i(K)=C_i(K,\mathbb{Z})$, is a vector space over the integer field $\mathbb{Z}$ with basis $S_i(K)$. The $i$-th \textit{cochain group} $C^i(K)=C^i(K,\mathbb{Z})$ is the dual of the chain group which can be defined by $C^i(K):=\text{Hom}(C_i(K),\mathbb{Z})$. Here Hom$(C_i,\mathbb{Z})$ is the set of all homomorphisms of $C_i$ into $\mathbb{Z}$. For an $(i+1)$-simplex $\sigma=[v_0,\dots,v_{i+1}]$, its \textit{coboundary operator}, $\delta_i:C^{i+1}(K) \rightarrow C^{i}(K)$, is defined as
$$
(\delta_i f)(\sigma)=\sum_{j=1}^{i+1}(-1)^j f([v_0,\dots,\hat{v}_j,\dots,v_{i+1}]),
$$
where $\hat{v}_j$ denotes that the vertex $v_j$ has been omitted. The \textit{boundary operators}, $\delta_i^*$, are the adjoints of the coboundary operators,
$$
\cdots C^{i+1}(K) \overset{\delta_i}{\underset{\delta_i^*}\rightleftarrows} C^i(K) \overset{\delta_{i-1}}{\underset{\delta_{i-1}^*}\rightleftarrows} C^{i-1}(K) \cdots
$$
satisfying $(\delta_i a,b)_{C^{i+1}} = (a,\delta_i^*b)_{C^i}$ for every $a \in C^i(K)$ and $b \in C^{i+1}(K)$, where $(\cdot ,\cdot)_{C^i}$ denote the scalar product on the cochain group. 

Lastly, we explain how tocreate a simplicial complex from an undirected graph, namely the \textit{clique complex}.

\begin{definition}
The clique complex $Cl(G)$ of an undirected graph $G$ is a simplicial complex where vertices of $G$ are its vertices and each $k$-clique, i.e., the complete subgraphs with $k$ vertices, in $G$ corresponds to a $(k-1)$-simplex in $Cl(G)$.
\end{definition}

\begin{figure*}[h!]
    \centering
    \begin{subfigure}[t]{0.5\textwidth}
        \centering
        \includegraphics[width=.8\textwidth]{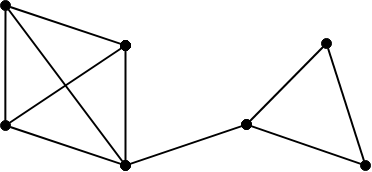}
        \caption{A graph $G$}
    \end{subfigure}%
    ~ 
    \begin{subfigure}[t]{0.5\textwidth}
        \centering
        \includegraphics[width=.8\textwidth]{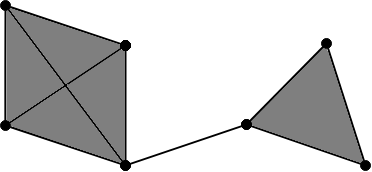}
        \caption{The clique complex $Cl(G)$ of the graph on the left}
    \end{subfigure}
    \caption{An example for constructing the clique complex of a graph (borrowed from~\cite{aktas2019persistence}).}
    \label{fig:clique}
\end{figure*}

For example, in Figure \ref{fig:clique}-a, there is a graph with a 4-clique on the left, 2-clique in the middle and 3-clique on the right. Hence, its clique complex, Figure \ref{fig:clique}-b, has a 3-simplex (tetrahedron), a 1-simplex (edge) and a 2-simplex (triangle). 

\subsubsection{Diffusion Fr{\'e}chet Function on Manifolds}\label{sec:dff}
Originally, Martinez et al. \cite{martinez2016multiscale,martinez2019probing} define the diffusion Fr{\'e}chet function (DFF) on manifolds, and particularly networks, to observe and quantify the local variations on data inspiring from the heat (diffusion) equation. Let $X$ be a compact $C^{\infty}$ manifold in $\mathbb{R}^n$. Recall the heat equation PDE
$$
\displaystyle \left( \frac{\partial}{\partial t} - \Delta_X\right ) u=0
$$
where the operator $\Delta_X$ is the Laplace-Beltrami operator, which is the infinitesimal generator of the heat diffusion process. $\Delta_X$ is a self-adjoint, positive semi-definite operator that acts on square integrable functions defined on $X$. Since $X$ is compact, there exists a countable orthonormal basis $\{\phi_1, \phi_2, \dots \}$ in the space of the square-integrable functions $\pazocal{L}^2(X)$ with real eigenvalues $0 \leq \lambda_1 \leq \lambda_2\leq \dots$ such that $\Delta_X\phi_i=\lambda_i\phi_i$.

To define DFF on $X$, they first embed each point on $X$ to $\pazocal{L}^2(X)$. For each $x \in X$, they associate the function $k_t(x,\cdot)=k_{t,x}$ where $k_t: X \times X \rightarrow \mathbb{R}$, is the fundamental solution of the heat equation which can be defined as 
\begin{equation}\label{eq:kernel}
k_t(x,y)=\sum_{i=1}^{\infty} e^{-t\lambda_i}\phi_i(x)\phi_i(y).
\end{equation}
This function is also called the \textit{heat kernel}. Here, $k_t(x,y)$ can be interpreted as the temperature at time $t$ at point $y$ when the heat source is at the point $x$ at time 0. The larger $k_t(x,y)$ means that heat diffuses fast from $x$ to $y$. This is a crucial fact since local geometrical affinities between different points on $X$ can be encoded through this function. For example, if $X$ is a graph with $x$ and $y$ being its vertices, DFF contains information on their connectivity in $X$. The heat kernel can also be seen as a function that describes the topological and geometrical similarity between points points in $X$. 

Next, they define the \textit{diffusion distance}, $d_t$, between two points in $X$ as the distance between their functions in $\pazocal{L}^2(X)$ with the usual metric of this space, i.e.
\begin{equation}\label{eq:dif}
d_t^2(x,y)=|| k_{t,x}-k_{t,y}||_{\pazocal{L}^2(X)}^2=\int_X (k_{t,x}(z)-k_{t,y}(z))^2 dz
\end{equation}
where $k_{t,x}(\cdot)=k_t(x,\cdot)$ for $x \in X$. The main idea here is that if heat diffuses in a similar way from points $x, y \in X$ to any other point $z \in X$, the functions $k_{t,x}$ and $k_{t,y}$ will be close in $\pazocal{L}^2(X)$, hence $x$ and $y$ are close in $X$. When we substitute Equation \ref{eq:kernel} in Equation \ref{eq:dif}, the diffusion distance can be written as 
$$
d_t^2(x,y)=\sum_{k=1}^n e^{-2\lambda_kt}(\phi_k(i) - \phi_k(j))^2.
$$
This distance is robust to noise \cite{coifman2006diffusion}, which is a crucial fact for using it in data mining problems. 

Finally, instead of using the Euclidean distance in the classical Fr{\'e}chet function, they use the diffusion distance and define the diffusion Fr{\'e}chet function on manifold as follows.

\begin{definition}
Let $\alpha$ be some probability measure defined on $X$. The diffusion Fr{\'e}chet function for manifolds is defined as
$$
V_{\alpha,t}:=\int_X d_t^2(x,y)\alpha(dy) = \int_X \sum_{k=1}^n e^{-2\lambda_kt}(\phi_k(i) - \phi_k(j))^2 \alpha(dy).
$$
\end{definition}

To define DFF on the vertices of a network, they first define a heat diffusion process on the network, and for this, they use the Laplacian on vertices. Let $G=(V, E)$ be an undirected, weighted network. The graph Laplacian matrix $\Delta$ is defined by 
$$
\Delta=D-W
$$
where $D$ is the diagonal degree matrix, and $W$ is the weighted adjacency matrix. Then the diffusion Fr{\'e}chet function on a network can be defined as follows.

\begin{definition}\label{def:dff} 
Let $\rho=[\rho_1,\dots, \rho_n]^T \in \mathbb{R}^n$ be a probability distribution on the vertex set $V$ of an undirected weighted network $G$. For $t>0$, the diffusion Fr{\'e}chet function on vertex $v_i$ is defined as 
$$
F_{\rho,t}(i)=\sum_{j=1}^n d_t^2(i,j)\rho_j
$$
with 
$$
d_t^2(i,j)=\sum_{k=1}^n e^{-2\lambda_kt}(\phi_k(i) - \phi_k(j))^2
$$
where  t is the heat diffusion time, $0\leq \lambda_1 \leq \dots \leq \lambda_n$ are the eigenvalues of the graph Laplacian $\Delta$ with orthonormal eigenvectors $\phi_1,\dots,\phi_n$.
\end{definition}

Diffusion distance takes manifolds' and networks' topological and geometrical structure into consideration. Hence, DFF values provide information on the relative topology of a vertex in the network. In this paper, we extend the diffusion distance to higher-order network structures, namely $p$-th simplices for any $p\geq 0$, beyond the vertices.

\subsection{Related work}

\subsubsection{Graph classification}
In literature, there are four main graph classification methods: graph isomorphism~\cite{zelinka1975certain}, graph edit distance~\cite{bunke2000recent,blumenthal2020comparing}, graph kernels~\cite{bastiangc,gkernel,borgwardt2005protein,shervashidze2009efficient,shervashidze2011weisfeiler,borgwardt2005shortest,gartner2003graph} and graph neural networks~\cite{errica2019fair,kernelnn,zhang2018end}. These methods basically measure the similarity between networks by comparing the three network structures in these networks: network topology, vertex weights/attributes, and edge weights/attributes~\cite{aggarwal2010managing}. 

In graph isomorphism, one can check whether there is a graph isomorphism or subgraph isomorphism between given two graphs, i.e., match the vertices and edges between these graphs or their subgraphs. Graph edit distances (GED) count graph edit operations, such as node and edge insertion/deletion, that are necessary to transform one graph to another. While GED is very popular for many graph mining problems, exact GED computation is NP-hard. Therefore, many different heuristic algorithms are proposed to compute approximate solutions~\cite{BORIA202019,zeng2009ged,blumenthal2020comparing,riesen2009approximate,kasparged14}. Among them, local search based algorithms~\cite{zeng2009ged,BORIA202019} provide the tightest upper bounds for GED. 

The graph \textit{kernels}~\cite{kriege2020survey,nikolentzos2019graph}, that have attracted a lot of attention during the last decade, are functions employed to handle the non-linear structure of graph data.  They are used to measure the similarity between pairs of graphs. One can define kernels using different graph structures such as random walks, shortest paths, and graphlets. Graph kernel methods have high complexity due to a pairwise similarity calculation, which makes it challenging to apply on today's large graphs. To address the challenges of graph kernels, \textit{Graph neural networks} have emerged in recent years. Recent methods~\cite{errica2019fair,ying2018hierarchical,morris2019weisfeiler} on graph neural networks combine node features and graph topology while extracting the features of the graph for classification. When graphs are noisy or specific sub-networks are essential for the classification, the attention mechanism is used on neural networks~\cite{lee2018graph}. 

More recently, persistent homology has been used for graph classification~\cite{aktas2019persistence,carriere2019perslay,zhao2019learning}. In these studies, persistent homology is employed to extract the topological features of graphs that persist across multiple scales. The basic idea here is to replace the vertices with a parametrized family of simplicial complexes and encode the change of the topological features (such as the number of connected components, holes, voids) of the simplicial complexes across different parameters~\cite{aktas2019persistence}. 

\subsubsection{Diffusion Fr{\'e}chet function in network mining}
There are different applications of DFF in network mining problems. The author in \cite{martinez2016multiscale} studies the co-occurrence networks of microbial communities to analyze the fecal microbiota transplantation in the treatment of C. difficile infection. Besides, in \cite{aktas2019classification} and \cite{aktas2019text}, the authors use DFF to classify music networks and text networks respectively. After defining DFF on networks, they use DFF values for classification. Furthermore, in \cite{keil2018topological}, the authors use the diffusion Fr{\'e}chet function for the attributed network clustering problem in social networks. 

\section{Methodology}\label{sec:method}
In this section, we introduce the main parts of our method. We first show how to create simplex-weighted super-graphs from vertex-labeled networks. In the second part, we present how we define the diffusion Fr{\'e}chet function on simplicial complexes. Finally, we outline our feature extraction and classification methods. 

\subsection{Simplex-weighted super-graphs}
In this paper, we assume that vertices of a network are \textit{labeled}, i.e., there is a label function $l: V \rightarrow L$ from the vertex set $V$ to a finite label set $L$. 

In order to reduce the graph size and create a unified feature list, we first represent a given vertex-labeled network $ G=(V,E,l)$ as a compressed \textit{super-graph} $\mathcal{G}=(\mathcal{V},\mathcal{E})$. It consists of a \textit{super-node} set, $\mathcal{V}$, and a \textit{super-edge} set, $\mathcal{E}$; each super-node $\nu \in \mathcal{V}$ represents a unique vertex label of the network, each super-edge $(\nu,\mu) \in \mathcal{E}$ between two super-nodes $(\nu$ and $\mu)$ represents the edge between a vertex with the label $\nu$ and a vertex with the label $\mu$ in the original network. Besides, we define weights on super-nodes and super-edges to reflect the frequency and co-occurrences of labels. We assign weights to super-nodes as the number of the corresponding label occurrences weights to super-edges as the co-occurrence frequency of the labels. The formal definition of our representation is as follows. 

\begin{definition}[Compressed super-graphs]\label{def:net}
Let $G=(V,E,l)$ be a vertex-labeled network with $l:V \rightarrow L$ being the label function and $L$ being the set of unique labels of $V$. Let $\mathcal{G}=(\mathcal{V},\mathcal{E}, W_{\mathcal{V}},Y_{\mathcal{E}})$ be the simplex-weighted super-graph of $G$ where $\mathcal{V}$ is the set of super-nodes, $\mathcal{E}$ is the set of super-edges, $W_V$ is the set of super-node weights and $Y_E$ is the set of super-edge weights. $\mathcal{G}$ is defined as follows
\begin{itemize}
    \item $\mathcal{V}=\{\nu_1, \dots,\nu_{n_L} | n_L \text{ is the size of } L\}$,
    \item $\mathcal{E}=\{e_1,\dots,e_{m} | e_k=(\nu_i,\nu_j) \text{ for } 1 \leq i, j \leq n_L, 1 \leq k \leq m \text{ where } m \text{ is the number} \\  \text{of different co-occurrences of labels in } G \}$,
    \item $W_{\mathcal{V}}=\{\omega_1,\dots,\omega_{n_L}| \omega_i \text{ is the frequency of the label }l_i \in L \text{ for } 1 \leq i \leq n_L \}$
    \item $Y_{\mathcal{E}}=\{y_1, \dots,y_{m} | y_k \text{ is the frequency of the co-occurrence} \text{ of the labels } l_i,l_{j} \\ \text{ for } 1 \leq i,j \leq n_L, 1 \leq k \leq m\}$,
\end{itemize}
\end{definition}

Furthermore, as the next step, we assign weights on simplices of the super-graph to record the co-occurrence frequency of higher-order structures on graphs. 

\begin{definition}[Simplex weights]
Let $S_p$ be the $p$-simplices of a super-graph $\mathcal{G}$, i.e., $S_p=\{\sigma_1,\dots, \sigma_{n_p} |  \sigma_i=[v_0, \dots, v_p] \in Cl(G) \text{ for } 1 \leq i \leq n_p \text{ where } n_p \text{ is the number of } \\ p \text{-cliques in } \mathcal{G} \text{ with } p>1\}$. We define a weight function $z:S_p \rightarrow \mathbb{Z}^+$ on $S_p$ such that for any $\sigma_i \in S_p$, $z(\sigma_i)=\text{ min} \{y_1, \dots, y_r\}$ with $r={p+1 \choose 2}$, where $y_j$'s are the weights of the edges in $\sigma_i$.
\end{definition}
The weights defined here will be used to define DFF on simplicial complexes in the next section. We close this section with an example of a simplex-weighted super-graph.

\begin{figure}[h!]
    \centering
    \begin{subfigure}[t]{0.68\textwidth}
        \centering
        \includegraphics[height=3in]{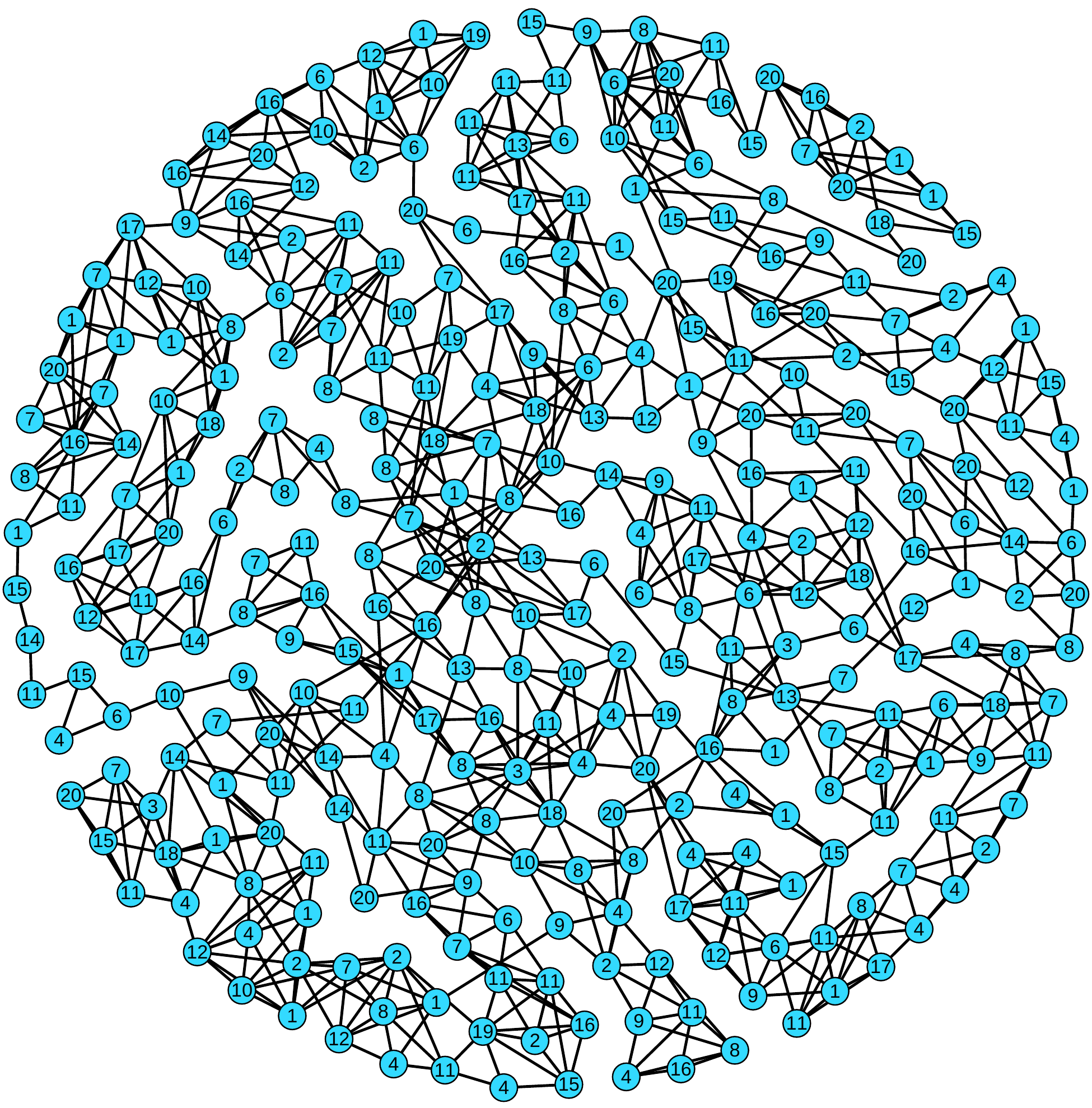}
        
        \caption{Original network $G$}
    \end{subfigure}%
    ~ 
    \begin{subfigure}[t]{0.3\textwidth}
        \centering
        \vspace{-5.2cm}
        \includegraphics[height=1.5in]{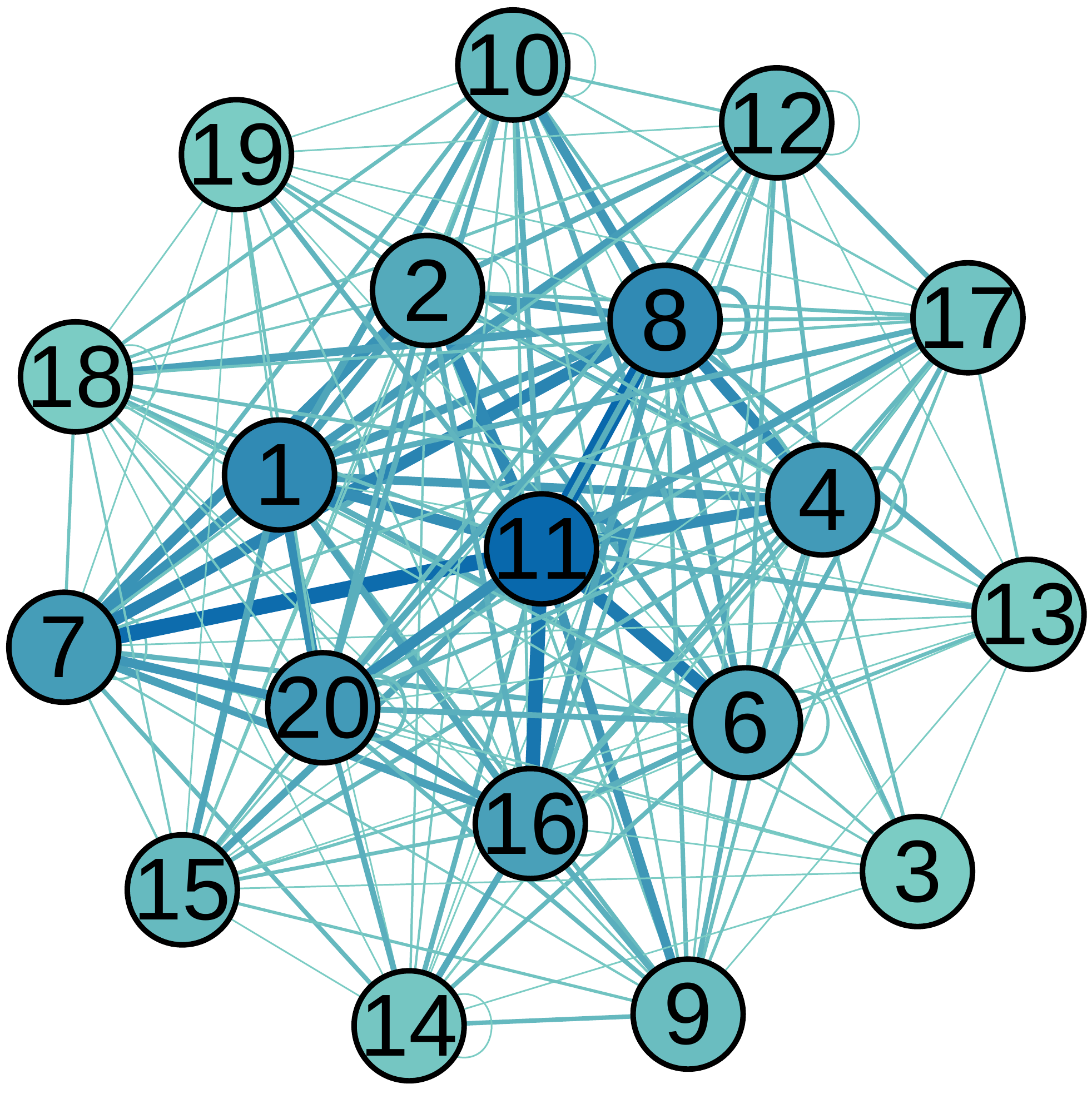}
        \vspace{0.78cm}
        \caption{Weighted super-graph $\mathcal{G}$ of $G$ in (a)}
    \end{subfigure}%
    ~ 
    \caption{Example of our simplex-weighted super-graph representation of a network in the DD dataset. The darker vertices and edges have larger weights in (b). }
    \label{fig:toycomp}
\end{figure}

\begin{example}
In Figure \ref{fig:toycomp}, we present a simplex-weighted super-graph of a vertex-labeled network in the DD dataset. While the original network has 327 vertices (0-simplices), 1798 edges (1-simplices), and 4490 triangles (2-simplices), its simplex-weighted super-graph has only 19 vertices, 172 edges and 1031 triangles since the original graph has only 19 different vertex labels. The darker vertices and edges have larger weights in Figure \ref{fig:toycomp}-(b). 
\end{example}

\subsection{Diffusion Fr{\'e}chet function on simplicial complexes}
In this section, we explain how to extend the diffusion Fr{\'e}chet function to any dimensional simplices beyond the vertices. For this, we first explain the simplicial Laplacian, and then we use it to define DFF on simplicial complexes.

\subsubsection{Simplicial Laplacian}
As we show in Section \ref{sec:dff}, the Laplacian matrix $\Delta$ is used to define the heat diffusion among the vertices of a graph through edges. However, as we explained before, this model is only considers the pairwise relations between vertices and ignores higher-order structures in graphs. In order to model heat diffusion on higher dimensional simplices and extract higher-order structures in graphs, we need to use an analog of the Laplacian matrix $\Delta$ for higher dimensions. In \cite{horak2013spectra}, the three \textit{simplicial Laplacian operators} for higher-dimensional simplices, using the boundary and coboundary operators between chain groups, are defined as
$$\pazocal{L}_p^{\text{down}}=\delta_{p-1}\delta_{p-1}^* \textit{ ~~~~~~~~~~ down Laplacian~~}$$
$$\pazocal{L}_p^{\text{up}}(K)=\delta_p^*\delta_p              \textit{ ~~~~~~~~~~~~~        up Laplacian~~~~~}$$
$$~ \pazocal{L}_p(K)=\pazocal{L}_p^{\text{up}}+\pazocal{L}_p^{\text{down}} \textit{   ~~~~~  Laplacian~~~~~~~~~~}$$
These operators are self-adjoint, non-negative, compact and have different spectral properties \cite{horak2013spectra}.

To make the expression of Laplacian explicit, they identify each coboundary operator $\delta_p$ with an incidence matrix $D_p$ in \cite{horak2013spectra}. The \textit{incidence matrix} $D_p \in \mathbb{Z}_2^{n_{p+1}} \times \mathbb{Z}_2^{n_p}$ encodes which $p$-simplices are incident to which $(p+1)$-simplices where $n_p$ is number of $p$-simplices. It is defined as
$$
D_p(i,j)= \left\{\begin{tabular}{ll}
 $1$ & if $\sigma_j^p$ is on the boundary of $\sigma_i^{p+1}$ \\
$0$  & otherwise
\end{tabular}\right.
$$
Here, we assume the simplices are not oriented. One can incorporate the orientations by simply adding ``$D_p(i,j) = -1$ if $\sigma_j^p$ is not coherent with the induced orientation of $\sigma_i^{p+1}$" in the definition if needed.

Furthermore, we assume that the simplices are weighted, i.e. there is a weight function $z$ defined on the set of all simplices of $K$ whose range is $\mathbb{R}^{+}$. let $W_p$ be an $n_p \times n_p$ diagonal matrix with $W_p(j,j)=z(\sigma_j^p)$ for all $j \in \{1,\dots, n_p\}$. Then, the $i$-dimensional up Laplacian can be expressed as the matrix
$$\pazocal{L}_i^{\text{up}}=W_i^{-1}D_i^{T}W_{i+1}D_i.$$
Similarly, the $i$-dimensional down Laplacian can be expressed as the matrix 
$$ \pazocal{L}_i^{\text{down}}=D_{i-1}W_{i-1}^{-1}D_{i-1}^{T}W_i.$$
Then, to express the $i$-dimensional Laplacian $\pazocal{L}_i$, we can add these two matrices. 

\subsubsection{Diffusion Fr{\'e}chet function on simplicial complexes}

The graph Laplacian used in Definition \ref{def:dff} is the $0$-th dimensional up Laplacian, i.e. $\Delta=D-W=\pazocal{L}_0^{\text{up}}$. Hence, in the case of DFF on vertices, we assume that the heat source is on vertices, and heat diffuses between vertices through edges. However, heat diffusion, the diffusion distance, and DFF can also be defined on $p$-simplices using the simplicial Laplacian operators for any $p>0$. In the case of down Laplacian, $\pazocal{L}_p^{\text{down}}$, we assume that heat sources are located on $p$-simplices and heat diffuses through $(p-1)$-simplices. In the case of up Laplacian, $\pazocal{L}_p^{\text{up}}$, we assume that heat sources are located on $p$-simplices and heat diffuses through $(p+1)$-simplices. In the case of Laplacian, $\pazocal{L}_p$, we assume that heat sources are located on $p$-simplices and heat diffuses both through $(p-1)$-simplices and $(p+1)$-simplices. 

 
Moreover, to define DFF on $p$-simplices with $p\geq 0$, we need to define a probability distribution on these simplices. We define a probability distribution on $p$-simplices using the simplex weights as follows.

\begin{definition}[Probability distribution on $p$-simplices]
Let $\mathcal{G}=(\mathcal{V},\mathcal{E},\\ W_\mathcal{V},Y_\mathcal{E})$ be a simplex-weighted super-graph and $S_p=\{\sigma_1, \dots , \sigma_k\}$ be its $p$-simplices with weights $z(S_p)=\{z_1, \dots ,z_k\}$. Then the probability distribution $\rho = [\rho_1, \dots ,\rho_k]^T \in \mathbb{R}^k$ is defined as
$$
\rho_i=\frac{z_i}{\sum_{j=1}^k z_j}
$$
for $1\leq i \leq k$.
\end{definition}

This definition allows us to make larger weights having more contribution to the diffusion Fr{\'e}chet function. This fact is crucial since the labels that occur more often in a network can have a vital role in classification.

Using Laplacian on simplicial complexes and probability distribution on $p$-simplices, we can define the diffusion Fr{\'e}chet  function on simplicial complexes as follows.

\begin{definition}[DFF on simplicial complexes]\label{def:dffsc} 
Let $\rho=[\rho_1,\dots, \rho_{n_p}]^T \in \mathbb{R}^{n_p}$ be a probability distribution on $p$-simplices of a simplicial complex $K$. For $t>0$, the diffusion Fr{\'e}chet function on a $p$-simplex $\sigma_i$ is defined as 
$$
F_{\rho,t}(i)=\sum_{j=1}^{n_p} d_t^2(i,j)\rho_j
$$
with 
$$
d_t^2(i,j)=\sum_{k=1}^{n_p} e^{-2\lambda_kt}(\phi_k(i) - \phi_k(j))^2
$$
where $0\leq \lambda_1 \leq \dots \leq \lambda_{n_p}$ are the eigenvalues of the Laplacian $\pazocal{L}_p$ with orthonormal eigenvectors $\phi_1,\dots,\phi_{n_p}$. 
\end{definition}
We can define the \textit{up diffusion Fr{\'e}chet function} $F_{\mathcal{E},t}^{\text{up}}$ and \textit{down diffusion Fr{\'e}chet function} $F_{\mathcal{E},t}^{\text{down}}$ by using the up Laplacian $\pazocal{L}_p^{\text{up}}$ and the down Laplacian $\pazocal{L}_p^{\text{down}}$ in the definition respectively. All these functions extract the topology and geometry of higher-order graph structures thanks to heat diffusion in different directions. 

\subsection{Feature extraction and classification}

For feature extraction, we first need to define the feature list for networks, and we employ the vertex labels for this sake. For 0-simplices, i.e., vertices, the feature list is the unique vertex labels observed in a given dataset. We first compute DFF on vertices. Then, we assign the reciprocal of that label's vertex' DFF value as the corresponding feature for that network. We use the reciprocal of the DFF values as features since the vertices of high degree with larger vertex and edge weights have smaller DFF values. However, these vertices are more critical for networks. Hence they need to have larger feature value to stress their importance. If a label in the feature list is not observed in a network, we assign zero as the corresponding feature. This process provides a feature vector for each network. 

For $p$-simplices with $p>0$, such as edges, triangles, we use vertex labels to label simplices and take the feature list as the all different labels of $p$-simplices observed in a given dataset. Then, we compute DFF on these $p$-simplices. As we do in the vertex case, if a label in the feature list is not observed in a network, we assign zero as the corresponding feature for the document. Otherwise, we assign the reciprocal of that label's simplex' DFF value as the corresponding feature for that network. This process provides a feature vector for each network for a fixed dimension of $p$. 

\begin{figure}[h!]
    \centering
\vspace{-1.5cm}
    \begin{subfigure}[t]{0.33\textwidth}
        \centering
        \includegraphics[height=1.6in]{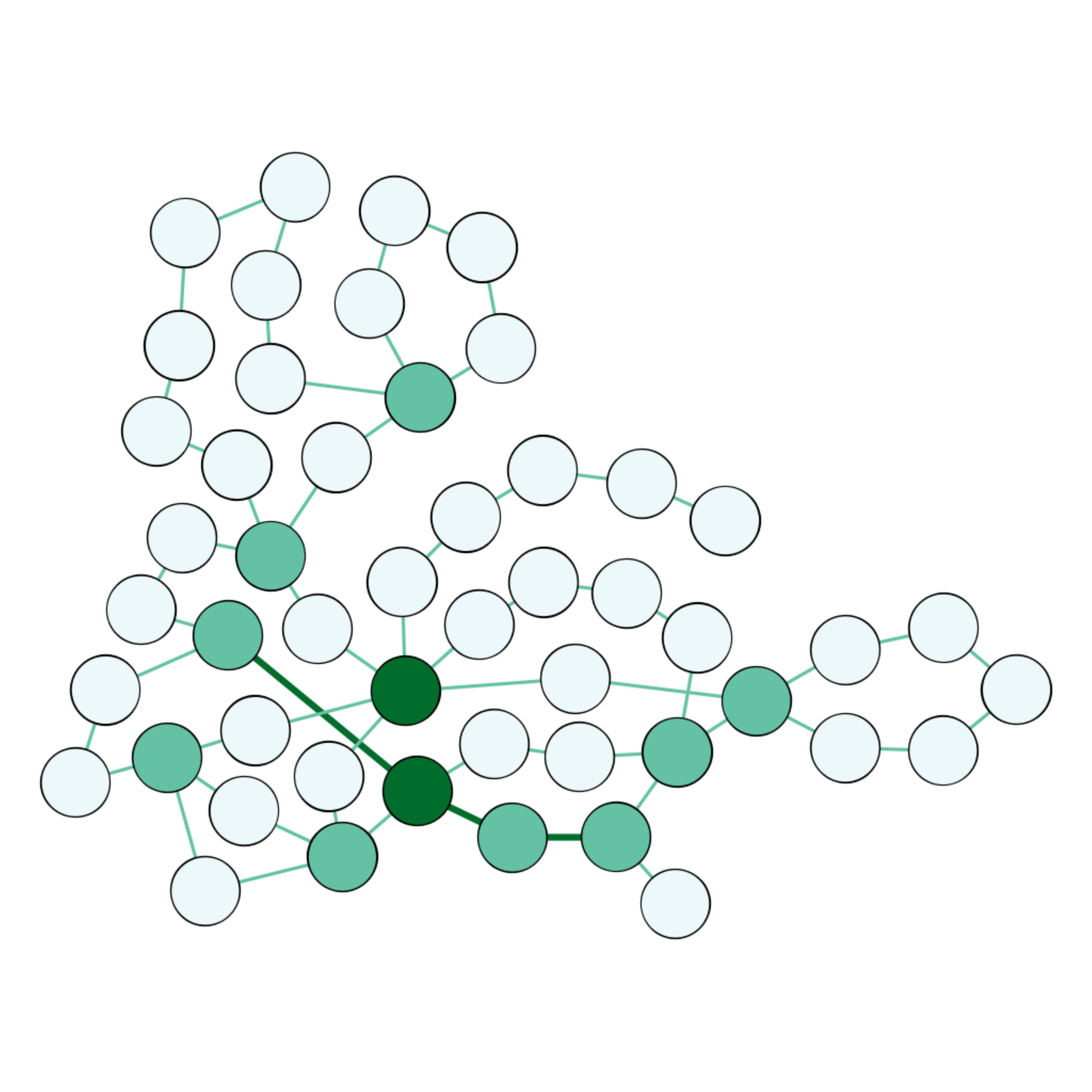}
        \caption{Original network}
        \label{fig:dff0}
    \end{subfigure}%
    ~ 
    \begin{subfigure}[t]{0.33\textwidth}
        \centering
        \includegraphics[height=1.6in]{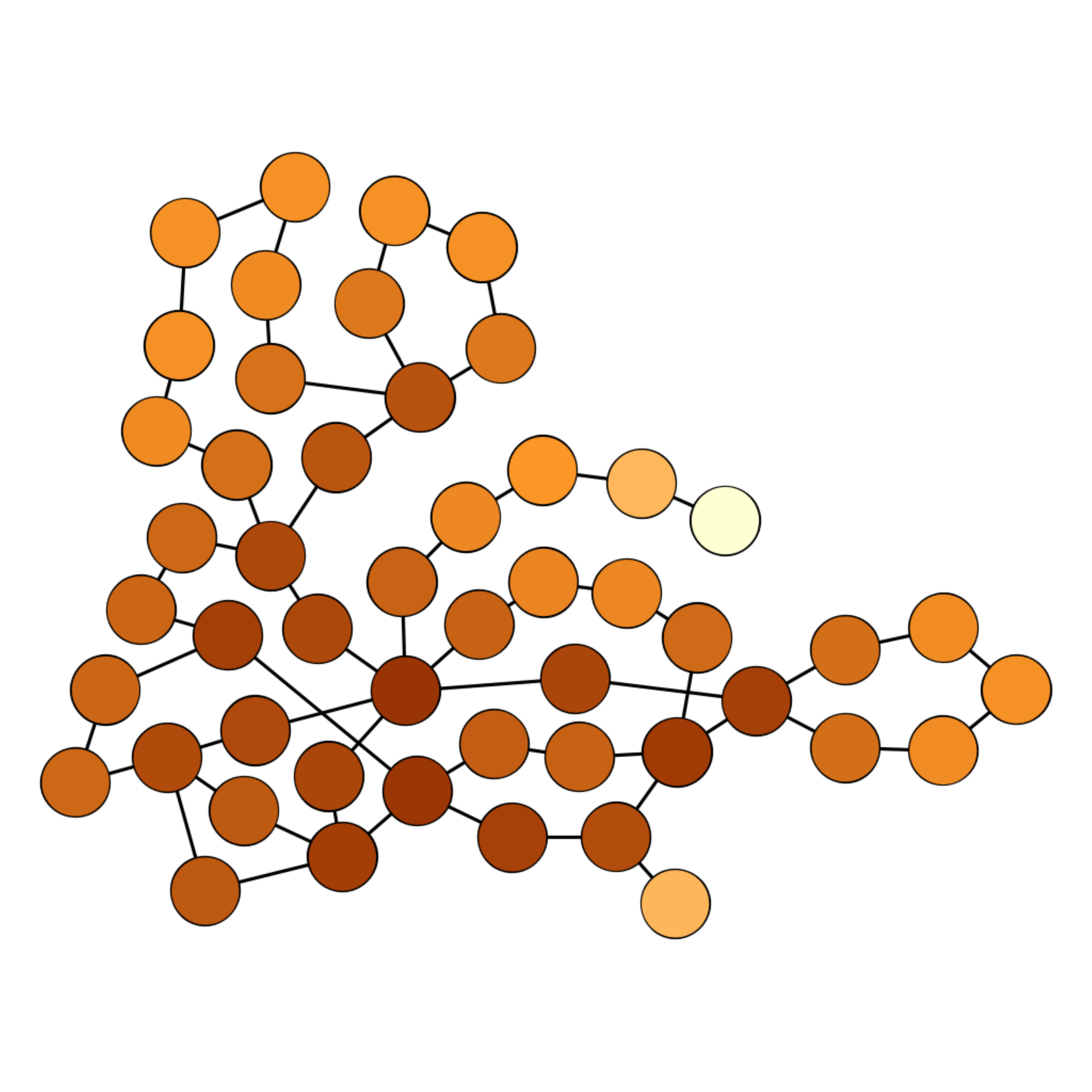}
        \caption{$t=1$}
        \label{fig:dff1}
    \end{subfigure}%
    ~ 
        \begin{subfigure}[t]{0.33\textwidth}
        \centering
        \includegraphics[height=1.6in]{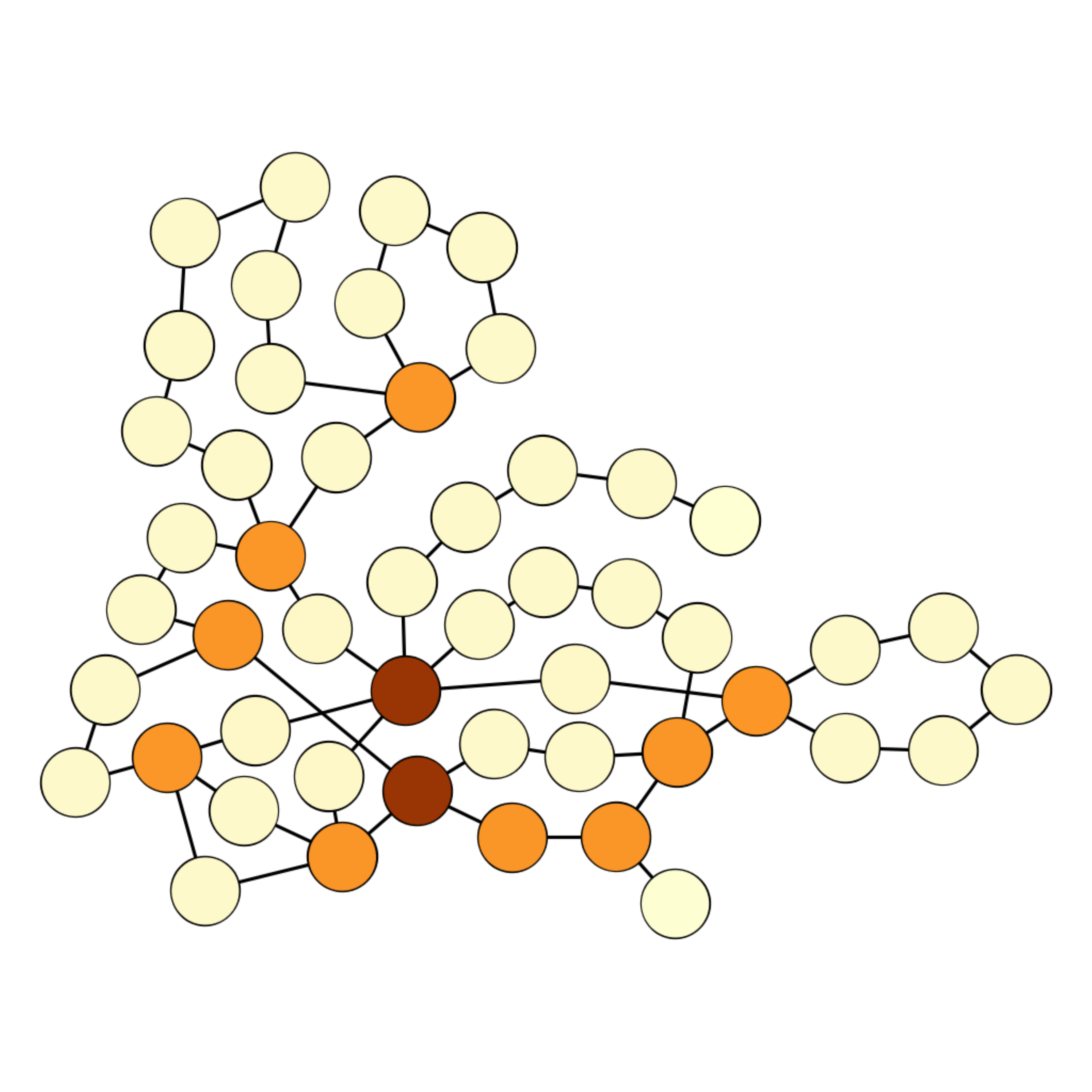}
        \caption{$t=10^{-3}$}
        \label{fig:dff2}
    \end{subfigure}%
    ~ 
    \caption{Diffusion Fr{\'e}chet function defined on a weighted network (a) in two different scale parameters (b) $t=1$, and (c) $t=10^{-3}$. While the darker vertices and edges have larger weights in (a), it means smaller diffusion Fr{\'e}chet function values in (b) and (c) (borrowed from \cite{aktas2019text}).}
    \label{fig:toynet}
\end{figure}

Furthermore, the scale parameter $t$ in Definition \ref{def:dff} and Definition \ref{def:dffsc}, which is the heat diffusion time as given by definitions, can be considered as the \textit{locality} index, i.e., the smaller the scale parameter value we use, the more local information on the networks we obtain. We can see the effect of this value in Figure \ref{fig:toynet}. In our experiments, we use different parameter values to understand their impact on classification.

After obtaining the features of each network up to a scale parameter value $t$ in the diffusion Fr{\'e}chet function, we train a classifier using these features of networks and their class values with a machine learning algorithm. 

\section{Experiments} \label{sec:exp}
In this section, we first introduce the datasets we use in our experiments. Then, we share the classification results for different simplex dimensions and scale parameter values in DFF. Lastly, we compare our method with the baseline graph classification methods.
\subsection{Datasets}
To test the efficiency of our method, we apply our framework to real-world bioinformatics networks: MUTAG, PROTEINS, PTC, and DD. MUTAG \cite{debnath1991structure} is a mutagenic nitro compound dataset, including 188 samples with two classes: aromatic and heteroaromatic. Its vertices have seven discrete labels. PROTEINS \cite{borgwardt2005protein} is a protein-protein interaction network whose vertices are secondary structure elements, and vertices that are neighbors in the amino-acid sequence of in 3D space are connected with an edge. Its vertices have 3 discrete labels, representing \textit{helix, sheet} or \textit{turn}. PTC \cite{toivonen2003statistical} is a dataset of 344 chemical compounds about the carcinogenicity for male and female rats, and its vertices have 19 discrete labels. DD \cite{dobson2003distinguishing} is a collection of 1,178 protein network structures with 82 discrete vertex labels, where each graph is classified as enzyme or non-enzyme class. Table \ref{table:datasets} has a summary of the statistics of these datasets.

\begin{table}[h!]
\centering
\large
\caption{Properties of the network datasets used in experiments.}
\begin{tabular}{|c|c|c|c|c|c|}

\hline   Datasets  & Size & Classes & Avg. nodes  & Avg. edges & Labels  \\
\hline MUTAG & 188 & 2 & 17.9 & 19.8 &	7\\	
\hline PROTEINS & 1113 & 2 & 39.1 & 72.9 &	3\\	
\hline PTC & 344 & 2 & 14.3 & 14.7 & 19 \\
\hline DD & 1178 & 2 & 284.3 & 715.7 & 82 \\ 
\hline

\end{tabular}
\label{table:datasets}
\end{table}

\subsection{Results}
In our experiments, we compute DFF on vertices, edges, and triangles (i.e., 0-, 1- and 2-simplices). For vertices, we define the heat diffusion only for up Laplacian since down Laplacian is not defined for vertices. For edges, we define the heat diffusion using down Laplacian, up Laplacian, and both. Lastly, we only define down Laplacian for triangles since we only need to compute the 1-dimensional incidence matrix $D_1$. Furthermore, we use the simplex weights of $p$-simplices with $p \in \{0,1,2\}$ only in the probability distribution, but not in Laplacian (i.e., we assume $W_p$ is the identity matrix in $\pazocal{L}_p$), to keep the Laplacian matrix symmetric. 

Besides, we use the parameter values $t \in \{1, 10^{-1}, 10^{-2}, 10^{-3}, 10^{-4}, 10^{-5} \}$ to test the efficiency of our method for different parameter values. 

After obtaining the feature vector of each network, we apply the Random Forest classification algorithm to build our prediction model. We use the 10-fold cross-validation process to evaluate our method. Finally, we obtain classification results (accuracy) for each dataset and different $t$ values in Table \ref{table:MUT}. 

\begin{table}[h!]
\centering
\large
\caption{Classification results (accuracy) for different simplex dimensions and scale parameter $t$ values. The cells for the best result in each column for each dataset is colored gray.}
\begin{tabular}{|c|c|c|c|c|c|c|c|}

\hline  & & $1$ & $10^{-1}$ & $10^{-2}$  & $10^{-3}$ & $10^{-4}$ & $10^{-5}$  \\ \hline
\hline \multirow{5}{*}{\rotatebox{90}{MUTAG}} & Vertex-up & 81.91 & 82.98 &	82.45 &	82.45 &	84.04 &	82.98\\	
\cline{2-8} & Edge-down &  \cellcolor{gray!35}85.64 &	 \cellcolor{gray!35}86.17 &	86.70 &	86.17 &	87.23 &	86.70\\	
\cline{2-8} & Edge-up & 82.45 & 84.04 & 86.70 & 86.17 & 87.23 &  \cellcolor{gray!35}88.30  \\
\cline{2-8} & Edge-both & 84.04 & 84.04 &  \cellcolor{gray!35}87.77 &	 \cellcolor{gray!35}88.30 &	 \cellcolor{gray!35}88.30 &	87.77 \\ 
\cline{2-8} & Triangle-down & 84.57 &	84.04 &	84.04 &	81.91 &	82.45 &	83.51   \\ 
\hline
\hline \multirow{5}{*}{\rotatebox{90}{PROTEINS}} & Vertex-up & 67.20 & 69.84 &  \cellcolor{gray!35}76.46 &  \cellcolor{gray!35}72.75 &  \cellcolor{gray!35}71.69 & 69.31\\	
\cline{2-8} &  Edge-down & 73.50 &	 \cellcolor{gray!35}73.23 &	69.81 & 69.90 &	70.17 &	70.89\\	
\cline{2-8} &  Edge-up & 72.42 & 72.42 & 74.39 & 71.25 & 69.90 & \cellcolor{gray!35}71.34 \\
\cline{2-8} &  Edge-both & 70.44 & 71.52 &74.39 & 70.89 & 70.89 & 70.44 \\
\cline{2-8} &  Triangle-down &  \cellcolor{gray!35}74.03 & 71.70 & 69.99 & 69.00 & 68.64 & 69.27   \\ 
\hline
\hline \multirow{5}{*}{\rotatebox{90}{PTC}} & Vertex-up & 55.52 & 58.14 &	62.79 &	59.88 &	59.59 &	59.59\\	
\cline{2-8} &  Edge-down &  \cellcolor{gray!35}61.92 &	 \cellcolor{gray!35}60.47 &	60.47 &	 \cellcolor{gray!35}62.50 &	 \cellcolor{gray!35}61.05 &	 \cellcolor{gray!35}61.05\\	
\cline{2-8} &  Edge-up & 56.98 & 59.88 & 62.79 & 58.43 & 59.01 & 56.98  \\
\cline{2-8} &  Edge-both & 61.34 &  \cellcolor{gray!35}60.47 &  \cellcolor{gray!35}63.66 &	59.01 &	59.59 &	58.43 \\ 
\cline{2-8} &  Triangle-down & 54.65 &	55.52 &	54.36 &	54.94 &	53.78 &	54.65  \\ 
\hline
\hline \multirow{5}{*}{\rotatebox{90}{DD}} & Vertex-up & 71.22 &	76.32 &	76.91 &	\cellcolor{gray!35}77.67 &	 \cellcolor{gray!35}79.03 & \cellcolor{gray!35}78.35\\	
\cline{2-8} &  Edge-down & 75.55 &	75.89 &	73.34 &	71.65 &	73.43 &	73.60\\	
\cline{2-8} &  Edge-up & 73.26 & 74.79 & 76.15 & 75.98 & 73.94 & 72.16 \\
\cline{2-8} &  Edge-both & 74.70 &	75.13 &	74.28 &	75.30 &	74.36 &	73.17 \\ 
\cline{2-8} &  Triangle-down &  \cellcolor{gray!35}77.67 &	 \cellcolor{gray!35}77.08 &	 \cellcolor{gray!35}77.08 &	76.40 &	76.66 &	76.91 \\ 
\hline

\end{tabular}
\label{table:MUT}
\end{table}

While the best accuracy results are obtained with the heat diffusion on vertices in PROTEINS and DD, the heat diffusion on edges provides the best results in MUTAG and PTC. Furthermore, the heat diffusion on triangles gives the best results in DD for three different $t$ values since the DD dataset has many triangles comparing to other datasets.

Furthermore, the effect of the scale parameter values also changes depending on both dataset and dimension. For example, while the smaller $t$ values increase the accuracy in MUTAG for edges, it decreases the accuracy in PROTEINS and DD for triangles.

\subsection{Comparison}
\begin{figure}[h!]
\centering
\includegraphics[width=\textwidth]{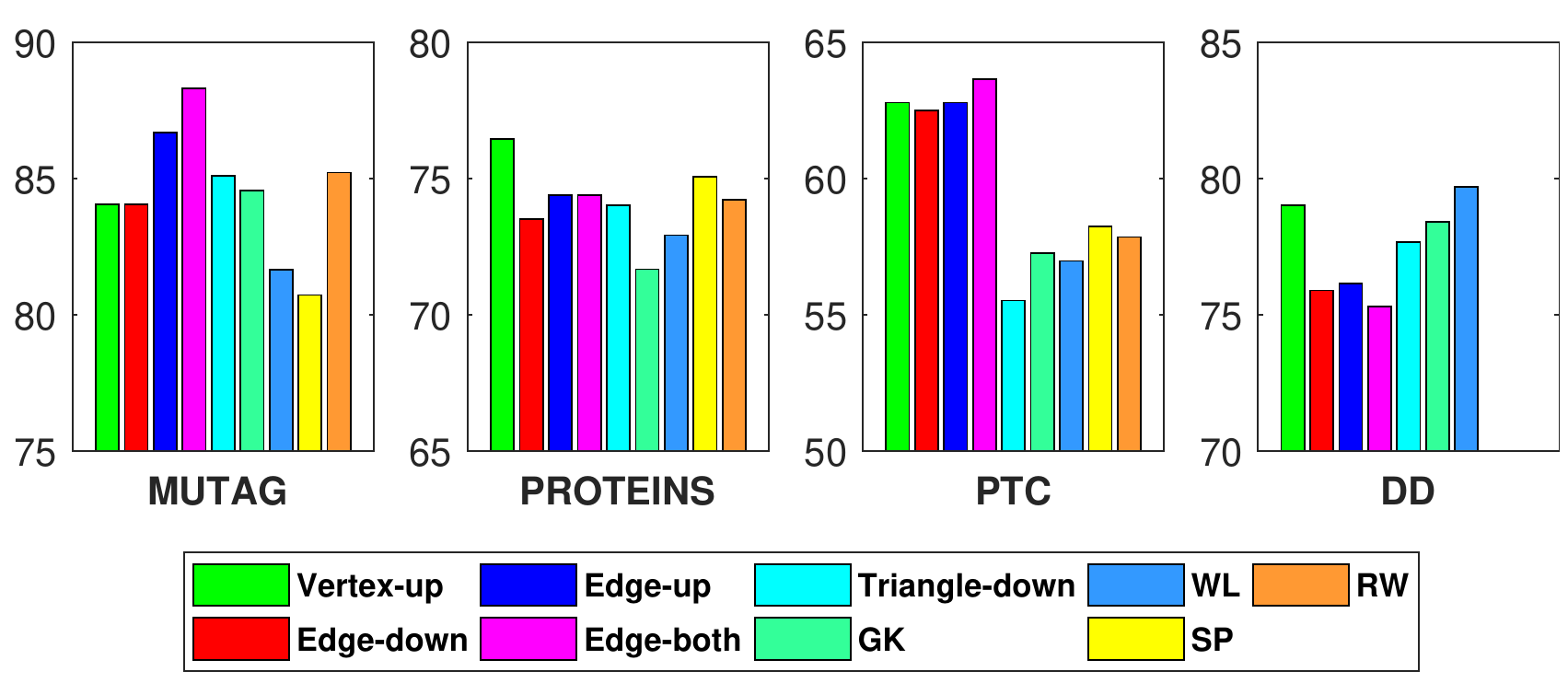}
\caption{Classification rates (accuracy) of the proposed algorithm and the baseline methods}
\label{fig:comp}
\end{figure}

We compare our model with the following baseline methods; graphlet kernel (GK) \cite{shervashidze2009efficient},  Weisfeiler-Lehman kernel (WL) \cite{shervashidze2011weisfeiler}, shortest-path kernel (SP) \cite{borgwardt2005shortest}, and random walk kernel (RW) \cite{gartner2003graph}. The same classification process has been applied with these methods for a fair comparison. The comparison results are in Table \ref{table:compare} and Figure \ref{fig:comp}. Here, for our each different method, we choose the $t$ values that gives the best results.

As we see in the table and the figure, our method performs better than the baseline methods on MUTAG, PROTEINS, and PTC and performs similar to WL in DD. This clearly indicates that higher-order network features obtained from heat diffusion on simplicial complexes is quite effective in the graph classification problem. 
\begin{table*}[h!]
\centering
\large
\caption{Classification rates (accuracy) of the proposed algorithms and the baseline methods. The cells for the best result in each column is colored gray. $>$ 24h indicates that the computation did not finish after 24 hours.}
\begin{tabular}{|c||c|c|c|c|}

\hline   & MUTAG & PROTEINS & PTC & DD  \\ \hline
\hline Vertex-up & 84.04 & \cellcolor{gray!35}76.46 & 62.79 & 79.03 \\
\hline Edge-down & 87.23 & 73.50 & 62.50 & 75.89 \\
\hline Edge-up & \cellcolor{gray!35}88.30 & 74.39 & 62.79 & 76.15 \\
\hline Edge-both & \cellcolor{gray!35}88.30 & 74.39 & \cellcolor{gray!35}63.66 & 75.30 \\
\hline Triangle-down & 84.57 & 74.03 & 55.52 & 77.67 \\ \hline
\hline GK & 81.66 & 71.67 & 57.26 & 78.40 \\
\hline WL & 80.72 & 72.92 & 56.97 & \cellcolor{gray!35}{79.70} \\
\hline SP & 85.22 & 75.07 & 58.24 & $>$ 24h \\
\hline RW & 83.72 & 74.22 & 57.85 & $>$ 24h \\
\hline
\end{tabular}
\label{table:compare}
\end{table*}

\section{Conclusion} \label{sec:conc}
In this paper, we propose a novel method for the graph classification problem. We first represent a vertex-labeled network with a simplex-weighted super-graph. We then define the diffusion Fr{\'e}chet function over the simplices of the networks to encode their non-trivial relations as the network features. We then classify graphs with the Random Forest classification algorithm and 10-fold cross-validation using these features. We show that the proposed algorithm has better classification results than the baseline methods. This paper is the first study in the literature showing the potential of using heat diffusion on simplicial Laplacians in the graph mining area. 

%
\section*{Conflict of interest}
The authors declare that they have no conflict of interest.

\bibliographystyle{spmpsci}      
\bibliography{mybib}   

\begin{thebibliography}{10}
\providecommand{\url}[1]{{#1}}
\providecommand{\urlprefix}{URL }
\expandafter\ifx\csname urlstyle\endcsname\relax
  \providecommand{\doi}[1]{DOI~\discretionary{}{}{}#1}\else
  \providecommand{\doi}{DOI~\discretionary{}{}{}\begingroup
  \urlstyle{rm}\Url}\fi

\bibitem{aggarwal2010managing}
Aggarwal, C.C., Wang, H.: Managing and mining graph data, vol.~40.
\newblock Springer (2010)

\bibitem{akbas2017attributed}
Akbas, E., Zhao, P.: Attributed graph clustering: An attribute-aware graph
  embedding approach.
\newblock In: Proceedings of the 2017 IEEE/ACM International Conference on
  Advances in Social Networks Analysis and Mining 2017, pp. 305--308 (2017)

\bibitem{akbas2017truss}
Akbas, E., Zhao, P.: Truss-based community search: a truss-equivalence based
  indexing approach.
\newblock Proceedings of the VLDB Endowment \textbf{10}(11), 1298--1309 (2017)

\bibitem{aktas2019text}
Aktas, M.E., Akbas, E.: Text classification via network topology: A case study
  on the holy quran.
\newblock In: 2019 18th IEEE International Conference On Machine Learning And
  Applications (ICMLA), pp. 1557--1562. IEEE (2019)

\bibitem{aktas2019persistence}
Aktas, M.E., Akbas, E., El~Fatmaoui, A.: Persistence homology of networks:
  methods and applications.
\newblock Applied Network Science \textbf{4}(1), 61 (2019)

\bibitem{aktas2019classification}
Aktas, M.E., Akbas, E., Papayik, J., Kovankaya, Y.: Classification of turkish
  makam music: a topological approach.
\newblock Journal of Mathematics and Music \textbf{13}(2), 135--149 (2019)

\bibitem{bao2017planning}
Bao, J., He, T., Ruan, S., Li, Y., Zheng, Y.: Planning bike lanes based on
  sharing-bikes' trajectories.
\newblock In: Proceedings of the 23rd ACM SIGKDD international conference on
  knowledge discovery and data mining, pp. 1377--1386 (2017)

\bibitem{battiston2020networks}
Battiston, F., Cencetti, G., Iacopini, I., Latora, V., Lucas, M., Patania, A.,
  Young, J.G., Petri, G.: Networks beyond pairwise interactions: structure and
  dynamics.
\newblock arXiv preprint arXiv:2006.01764  (2020)

\bibitem{blumenthal2020comparing}
Blumenthal, D.B., Boria, N., Gamper, J., Bougleux, S., Brun, L.: Comparing
  heuristics for graph edit distance computation.
\newblock The VLDB Journal \textbf{29}(1), 419--458 (2020)

\bibitem{borgwardt2005shortest}
Borgwardt, K.M., Kriegel, H.P.: Shortest-path kernels on graphs.
\newblock In: Fifth IEEE International Conference on Data Mining (ICDM'05), pp.
  8--pp. IEEE (2005)

\bibitem{borgwardt2005protein}
Borgwardt, K.M., Ong, C.S., Sch{\"o}nauer, S., Vishwanathan, S., Smola, A.J.,
  Kriegel, H.P.: Protein function prediction via graph kernels.
\newblock Bioinformatics \textbf{21}(suppl\_1), i47--i56 (2005)

\bibitem{BORIA202019}
Boria, N., Blumenthal, D.B., Bougleux, S., Brun, L.: Improved local search for
  graph edit distance.
\newblock Pattern Recognition Letters \textbf{129}, 19 -- 25 (2020)

\bibitem{bunke2000recent}
Bunke, H.: Recent developments in graph matching.
\newblock In: Proceedings 15th International Conference on Pattern Recognition.
  ICPR-2000, vol.~2, pp. 117--124. IEEE (2000)

\bibitem{carriere2019perslay}
Carriere, M., Chazal, F., Ike, Y., Lacombe, T., Royer, M., Umeda, Y.: Perslay:
  A neural network layer for persistence diagrams and new graph topological
  signatures.
\newblock stat \textbf{1050}, 17 (2019)

\bibitem{coifman2006diffusion}
Coifman, R.R., Lafon, S.: Diffusion maps.
\newblock Applied and computational harmonic analysis \textbf{21}(1), 5--30
  (2006)

\bibitem{Cook2006}
Cook, D.J., Holder, L.B.: Mining Graph Data.
\newblock John Wiley \& Sons (2006)

\bibitem{debnath1991structure}
Debnath, A.K., Lopez~de Compadre, R.L., Debnath, G., Shusterman, A.J., Hansch,
  C.: Structure-activity relationship of mutagenic aromatic and heteroaromatic
  nitro compounds. correlation with molecular orbital energies and
  hydrophobicity.
\newblock Journal of medicinal chemistry \textbf{34}(2), 786--797 (1991)

\bibitem{dobson2003distinguishing}
Dobson, P.D., Doig, A.J.: Distinguishing enzyme structures from non-enzymes
  without alignments.
\newblock Journal of molecular biology \textbf{330}(4), 771--783 (2003)

\bibitem{duvenaud2015convolutional}
Duvenaud, D.K., Maclaurin, D., Iparraguirre, J., Bombarell, R., Hirzel, T.,
  Aspuru-Guzik, A., Adams, R.P.: Convolutional networks on graphs for learning
  molecular fingerprints.
\newblock In: Advances in neural information processing systems, pp. 2224--2232
  (2015)

\bibitem{errica2019fair}
Errica, F., Podda, M., Bacciu, D., Micheli, A.: A fair comparison of graph
  neural networks for graph classification.
\newblock arXiv preprint arXiv:1912.09893  (2019)

\bibitem{gartner2003graph}
G{\"a}rtner, T., Flach, P., Wrobel, S.: On graph kernels: Hardness results and
  efficient alternatives.
\newblock In: Learning theory and kernel machines, pp. 129--143. Springer
  (2003)

\bibitem{horak2013spectra}
Horak, D., Jost, J.: Spectra of combinatorial laplace operators on simplicial
  complexes.
\newblock Advances in Mathematics \textbf{244}, 303--336 (2013)

\bibitem{keil2018topological}
Keil, W., Aktas, M.: Topological data analysis of attribute networks using
  diffusion frechet function with ego-networks.
\newblock In: The 7th International Conference on Complex Networks and Their
  Applications (extended Abstract), Cambridge, United Kingdom, pp. 194--196
  (2018)

\bibitem{kriege2020survey}
Kriege, N.M., Johansson, F.D., Morris, C.: A survey on graph kernels.
\newblock Applied Network Science \textbf{5}(1), 1--42 (2020)

\bibitem{lee2018graph}
Lee, J.B., Rossi, R., Kong, X.: Graph classification using structural
  attention.
\newblock In: Proceedings of the 24th ACM SIGKDD International Conference on
  Knowledge Discovery \& Data Mining, pp. 1666--1674 (2018)

\bibitem{martinez2016multiscale}
Martinez, D.H.D.: Multiscale summaries of probability measures with
  applications to plant and microbiome data.
\newblock Ph.D. thesis, The Florida State University (2016)

\bibitem{martinez2019probing}
Mart{\'\i}nez, D.H.D., Lee, C.H., Kim, P.T., Mio, W.: Probing the geometry of
  data with diffusion fr{\'e}chet functions.
\newblock Applied and Computational Harmonic Analysis \textbf{47}(3), 935--947
  (2019)

\bibitem{morris2019weisfeiler}
Morris, C., Ritzert, M., Fey, M., Hamilton, W.L., Lenssen, J.E., Rattan, G.,
  Grohe, M.: Weisfeiler and leman go neural: Higher-order graph neural
  networks.
\newblock In: Proceedings of the AAAI Conference on Artificial Intelligence,
  vol.~33, pp. 4602--4609 (2019)

\bibitem{Newman06062006}
Newman, M.J.: Modularity and community structure in networks.
\newblock Proceedings of the National Academy of Sciences \textbf{103}(23),
  8577--8582 (2006).
\newblock National Academy of Sciences

\bibitem{kernelnn}
Nikolentzos, G., Meladianos, P., Tixier, A.J.P., Skianis, K., Vazirgiannis, M.:
  Kernel graph convolutional neural networks.
\newblock In: V.~K{\r{u}}rkov{\'a}, Y.~Manolopoulos, B.~Hammer, L.~Iliadis,
  I.~Maglogiannis (eds.) Artificial Neural Networks and Machine Learning --
  ICANN 2018, pp. 22--32. Springer International Publishing, Cham (2018)

\bibitem{nikolentzos2019graph}
Nikolentzos, G., Siglidis, G., Vazirgiannis, M.: Graph kernels: A survey.
\newblock arXiv preprint arXiv:1904.12218  (2019)

\bibitem{przulj2003graph}
Przulj, N.: Graph theory approaches to protein interaction data analysis.
\newblock Tech. rep., University of Toronto (2004)

\bibitem{bastiangc}
Rieck, B., Bock, C., Borgwardt, K.: A persistent weisfeiler-lehman procedure
  for graph classification.
\newblock In: K.~Chaudhuri, R.~Salakhutdinov (eds.) Proceedings of the 36th
  International Conference on Machine Learning, \emph{Proceedings of Machine
  Learning Research}, vol.~97, pp. 5448--5458. PMLR, Long Beach, California,
  USA (2019)

\bibitem{riesen2009approximate}
Riesen, K., Bunke, H.: Approximate graph edit distance computation by means of
  bipartite graph matching.
\newblock Image and Vision computing \textbf{27}(7), 950--959 (2009)

\bibitem{kasparged14}
Riesen, K., Fischer, A., Bunke, H.: Combining bipartite graph matching and beam
  search for graph edit distance approximation.
\newblock In: N.~El~Gayar, F.~Schwenker, C.~Suen (eds.) Artificial Neural
  Networks in Pattern Recognition, pp. 117--128. Springer International
  Publishing, Cham (2014)

\bibitem{shervashidze2011weisfeiler}
Shervashidze, N., Schweitzer, P., Van~Leeuwen, E.J., Mehlhorn, K., Borgwardt,
  K.M.: Weisfeiler-lehman graph kernels.
\newblock Journal of Machine Learning Research \textbf{12}(77), 2539--2561
  (2011)

\bibitem{shervashidze2009efficient}
Shervashidze, N., Vishwanathan, S., Petri, T., Mehlhorn, K., Borgwardt, K.:
  Efficient graphlet kernels for large graph comparison.
\newblock In: Artificial Intelligence and Statistics, pp. 488--495 (2009)

\bibitem{tanner2019paper}
Tanner, W., Akbas, E., Hasan, M.: Paper recommendation based on citation
  relation.
\newblock In: 2019 IEEE International Conference on Big Data (Big Data), pp.
  3053--3059. IEEE (2019)

\bibitem{toivonen2003statistical}
Toivonen, H., Srinivasan, A., King, R.D., Kramer, S., Helma, C.: Statistical
  evaluation of the predictive toxicology challenge 2000--2001.
\newblock Bioinformatics \textbf{19}(10), 1183--1193 (2003)

\bibitem{gkernel}
Yanardag, P., Vishwanathan, S.: Deep graph kernels.
\newblock In: Proceedings of the 21th ACM SIGKDD International Conference on
  Knowledge Discovery and Data Mining, KDD ’15, p. 1365–1374. Association
  for Computing Machinery, New York, NY, USA (2015)

\bibitem{ying2018hierarchical}
Ying, Z., You, J., Morris, C., Ren, X., Hamilton, W., Leskovec, J.:
  Hierarchical graph representation learning with differentiable pooling.
\newblock In: Advances in neural information processing systems, pp. 4800--4810
  (2018)

\bibitem{zelinka1975certain}
Zelinka, B.: On a certain distance between isomorphism classes of graphs.
\newblock {\v{C}}asopis pro p{\v{e}}stov{\'a}n{\'\i} matematiky
  \textbf{100}(4), 371--373 (1975)

\bibitem{zeng2009ged}
Zeng, Z., Tung, A.K.H., Wang, J., Feng, J., Zhou, L.: Comparing stars: On
  approximating graph edit distance.
\newblock Proc. VLDB Endow. \textbf{2}(1), 25–36 (2009)

\bibitem{zhang2018end}
Zhang, M., Cui, Z., Neumann, M., Chen, Y.: An end-to-end deep learning
  architecture for graph classification.
\newblock In: Thirty-Second AAAI Conference on Artificial Intelligence (2018)

\bibitem{zhao2019learning}
Zhao, Q., Wang, Y.: Learning metrics for persistence-based summaries and
  applications for graph classification.
\newblock In: Advances in Neural Information Processing Systems, pp. 9855--9866
  (2019)

\end{thebibliography}

%
%

\end{document}